\documentclass[fleqn,12pt,twoside]{article}
\usepackage{espcrc1}
\usepackage{graphicx}
\usepackage{epsfig}
\usepackage[figuresright]{rotating}

\newcommand{\AmS}{{\protect\the\textfont2
  A\kern-.1667em\lower.5ex\hbox{M}\kern-.125emS}}

\title{Studies on the electromagnetic structure of the nucleon by free
and quasi-free Compton scattering at MAMI (Mainz)}
\author{Martin Schumacher\\
{Zweites Physikalisches Institut der
Universit\"at G\"ottingen\\
Bunsenstrasse 7-9, D-37073 G\"ottingen, Germany}
\thanks{Supported by Deutsche Forschungsgemeinschaft SPP(1034) and projects
  SCHU222 and 436 RUS 113/510. Presented on PANIC Osaka/Japan 
30.09.--04.10.2002, e-mail: mschuma3@gwdg.de}}
      
\begin{document}
\maketitle

\begin{abstract}
Using hydrogen and deuterium targets, Compton scattering by the proton 
and neutron have been studied at the tagged photon beam of the 
MAMI (Mainz) accelerator using  different experimental setups. 
The theoretical tools for the analysis
of the  experimental data have been investigated, as there are
the nonsubtracted dispersion theory and the theory of quasi-free
reactions on the proton and neutron bound in the deuteron. Experimental
Compton scattering data are  understood in the 
first and second resonance region with good precision. 
Precise electromagnetic polarizabilities and 
spin polarizabilities for the proton and the neutron have been determined. 

\end{abstract}

\section{COMPTON SCATTERING AND DEGREES OF FREEDOM}

For a long time it has been believed that electromagnetic polarizabilities
may be calculated using the quantum mechanical relations
$\alpha=2\sum_{n\neq 0}\frac{|\langle n^{(i)}|D_z|0\rangle|^2}{
E^{(i)}_n-E^{(i)}_0}+Z^2\frac{e^2\langle r^2_E\rangle}{
3M}, 
\beta=2\sum_{n\neq 0}\frac{|\langle n^{(i)}|M_z|0\rangle|^2}{
E^{(i)}_n-E^{(i)}_0} -e^2\sum_i\frac{q^2_i}{6m_i}\langle 0|\rho^2_i|0\rangle
-\frac{\langle|{\bf D}^2|0\rangle}{2M}.$
Recently, it has been shown that these relations  contain large uncertainties,
especially in the $r^2_E$ dependent  retardation term 
which is compensated
by other relativistic terms of the same order \cite{lee}. 
Problems
of an other kind are experienced when applying chiral perturbation theory.
Agreement with experimental data is  obtained when 
the calculation is restricted to the leading order.
Then the following relation  is obtained \cite{bernard95}
$\alpha_p=\alpha_n=10\beta_p=10\beta_n=\frac{5}{96\pi}
\left(\frac{g_A}{f_\pi}\right)^2\frac{e^2/4\pi}{m_\pi}\simeq12.6\cdot
10^{-4}{\rm fm}^3.$
However, the unavoidable inclusion of the $\Delta$ degree of freedom  
removes  the agreement  \cite{gries02}.

The two approaches also have in common that only a limited selection of
degrees of freedom is  taken into account and that it is difficult 
to identify degrees of freedom which are possibly missing. 
This problem is solved by  applying the nonsubtracted dispersion theory, 
starting with the following consideration:
In  the extreme forward ($\theta=0$) and extreme backward ($\theta=\pi$)
direction  the amplitudes for Compton scattering
may be written in the form
\cite{lvov97,babusci98}  
\begin{eqnarray}
&&T^{\rm LAB}(\theta=0)=f_0(\omega){\epsilon}'\cdot{\epsilon}+
g_0(\omega)\, \mbox{i}\, {\sigma}\cdot({\epsilon}'\times
{\epsilon})\\
&&T^{\rm LAB}(\theta=\pi)=f_\pi(\omega){\epsilon}'\cdot{\epsilon}+
g_\pi(\omega)\, \mbox{i}\,{\sigma}\cdot({{\epsilon}}'
\times
{{\epsilon}}).
\label{T2}
\end{eqnarray}
where
$f_{0,\pi}(\omega)= 1/2 \{T_{1/2}+T_{3/2}\}$ 
correspond to the case where the initial-state and final-state photons 
have parallel linear polarizations  and 
$g_{0,\pi}(\omega)= 1/2 \{T_{1/2}-T_{3/2}\}$ to the case where these 
photons have perpendicular linear polarizations. The relations between 
the amplitudes $f$ and $g$ and the invariant (LPS) \cite{lvov97,babusci98}
amplitudes $A_i$ are
\begin{footnotesize}
\begin{eqnarray}
&&f_0(\omega)= -\frac{\omega^2}{2\pi}\left[A_3(\nu,t)+ A_6(\nu,t)
\right],\quad\quad\quad
g_0(\omega)=\frac{\omega^3}{2\pi m}A_4(\nu,t), \\
&&f_\pi(\omega)=-\frac{\omega\omega'}{2\pi}\left(1+\frac{\omega\omega'}
{m^2}\right)^{1/2}\left[ 
A_1(\nu,t) + \frac{\omega\omega'}{m^2}A_5(\nu,t)\right],\\
&&g_\pi(\omega)=-\frac{\omega\omega'}{2\pi}\left(1+\frac{\omega\omega'}
{m^2}\right)^{-1/2}\frac{\omega+\omega'}{2m}\left[
A_2(\nu,t)+ \left(1+\frac{\omega\omega'}{m^2}\right)A_5(\nu,t)\right]
,\\
&&\omega'(\theta=\pi)=\frac{\omega}{1+2\frac{\omega}{m}},\,\,
\nu=\frac12 (\omega+\omega'),\,\, t(\theta=0)=0,
\,\,t(\theta=\pi)=-4\omega\omega.'
\label{T3}
\end{eqnarray}
\end{footnotesize}
For the electric, $\alpha$, and magnetic, $\beta$,  polarizabilities 
and the spin polarizabilities $\gamma_0$ and $\gamma_\pi$ for the
forward and backward directions, respectively, 
we obtain the relations
\begin{footnotesize}
\begin{eqnarray}
&&\alpha+\beta = -\frac{1}{2\pi}\left[A^{\rm nB}_3(0,0)+ 
A^{\rm nB}_6(0,0)\right], \quad 
\alpha-\beta = -\frac{1}{2\pi}
\left[A^{\rm nB}_1(0,0)\right], \nonumber\\ 
&&\gamma_0= \frac{1}{2\pi m}\left[A^{\rm nB}_4(0,0)
\right], \quad\quad\quad\quad \quad\quad\quad\,\,\,
\gamma_\pi = -\frac{1}{2\pi m}
\left[A^{\rm nB}_2(0,0)+A^{\rm nB}_5(0,0) \right].
\label{T4}
\end{eqnarray}
\end{footnotesize}
The amplitudes $A^{\rm nB}_i(s,t)$ are the non-Born parts of the
invariant (LPS) amplitudes,  with  the amplitudes
$A^{\rm nB}_1(s,t)$ and $A^{\rm nB}_2(s,t)$ being  dominated
by the $t-$channel contributions:
\begin{footnotesize}
\begin{equation}
\quad\quad A^{\sigma}_1(t)=\frac{g_{\sigma NN}
F\sigma\gamma\gamma}{t-m^2_\sigma},
\quad\quad
A^{{\pi^0}+\eta+\eta'}_2(t)=\frac{g_{{\pi^0} NN}F_{\pi^0\gamma\gamma}}
{t-m^2_{\pi^0}}\tau_3+\frac{g_{{\eta} NN}F_{\eta\gamma\gamma}}
{t-m^2_{\eta}}+\frac{g_{{\eta'} NN}F_{\eta'\gamma\gamma}}
{t-m^2_{\eta'}}.
\end{equation}
\end{footnotesize}
The structure of the $t-$channel part of $A_2$ seems to be well
understood in terms of the $\pi$, the $\eta$ and the $\eta'$ meson which 
couple to two photons with perpendicular polarization.
For the $t-$channel part of $A_1$ it tentatively has been suggested
that the $\sigma$ meson is the relevant particle which couples to two photons
with parallel polarization. This suggestion implies that  the $\sigma$ 
meson degree of freedom makes the most 
important contribution to $\alpha-\beta$ and, therefore, is responsible
for the diamagnetism and the enhancement of the electric polarizability
$\alpha$  over the s-channel prediction. An experimental proof
\cite{galler01,wolf01} for this suggestion  is shown in Fig. 1.
\begin{figure}[htb]
\begin{minipage}[b]{80mm}
\includegraphics[scale=0.4]{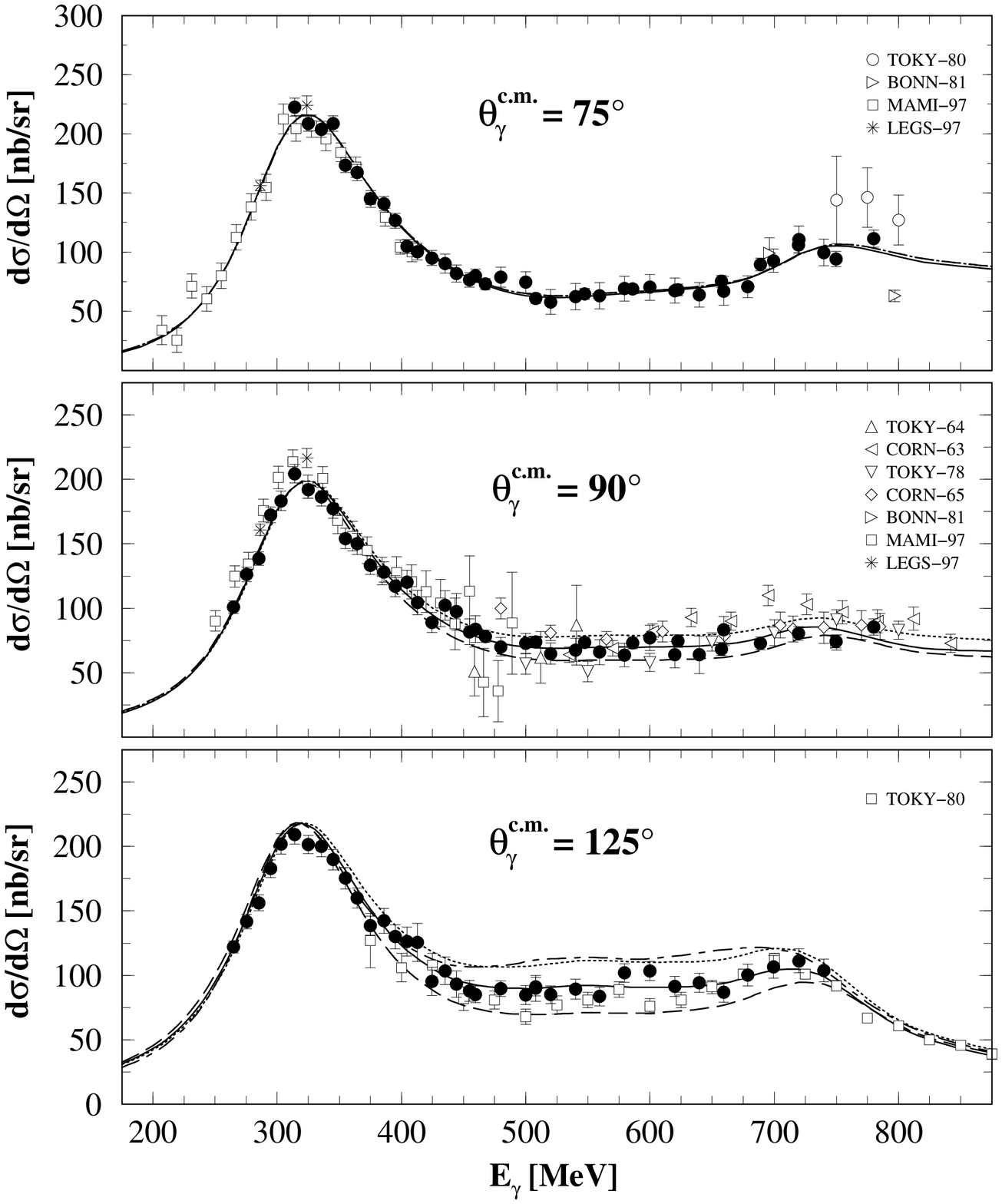}
\end{minipage}
\begin{minipage}[b]{70mm}
Fig. 1: Differential cross sections for Compton scattering by the proton
versus photon energy \cite{galler01,wolf01}. The three 
panels contain data corresponding to the
cm-angles of 75$^\circ$, 90$^\circ$ and  125$^\circ$. The three curves
are calculated for different mass parameters $m_\sigma=800$ MeV (upper), $600$
MeV (center) and $400$ MeV (lower). At large angles a good sensitivity
to $m_\sigma$ is observed with an optimum fit for $m_\sigma=600$ MeV.
This agreement may tentatively be considered as a confirmation of the 
$\sigma$ meson ansatz for the $t$-channel contribution to the invariant
scattering amplitude $A_1$.
\end{minipage}
\label{fig:fig9}
\end{figure}

\section{RECENT EXPERIMENTS ON THE POLARIZABILITIES OF PROTON AND NEUTRON}

\begin{table}
\caption{Structure constants for the proton and the neutron. The errors
attributed to $\gamma_\pi$ reflect the precision with which these
quantities are confirmed by experiments \cite{camen02,kossert02}. Sum-rule
predictions for $\gamma_\pi$ are discussed in \cite{lvov99}.} 
\label{table}
\begin{center}
\begin{tabular}{lll}
\hline
structure constant&proton&neutron\\
\hline
$\alpha$&$12.2 \pm 0.6$ & $12.5\pm 2.3$\\
$\beta$&$ 1.8\mp 0.6$& $2.7\mp 2.3$\\
$\gamma_\pi $&$-38.7\pm 1.8$&$+58.6\pm 4.0$\\
$\gamma_\pi(t-{\rm{channel}})$&$-46.6$&$+43.4$\\
$\gamma_\pi( s-{\rm{channel}})$&$+7.9\pm 1.8$&$+15.2\pm 4.0$\\
\hline
\end{tabular}
\end{center}
\end{table}

A new precise determination of the polarizability difference 
for the proton $\alpha_p-\beta_p$  has been performed using the 
TAPS apparatus 
at photon energies below pion threshold \cite{olmos01}. At these energies the 
contribution of the backward spin-polarizability $\gamma^{(p)}_\pi$
 to the scattering
amplitude is small so that $\alpha_p-\beta_p$ can be extracted from the 
data without mayor model dependencies. The result obtained is
\begin{equation}
\quad\quad
\alpha_p-\beta_p=10.5\pm 0.9({\rm stat+syst})\pm 0.7({\rm model})\cdot
10^{-4}{\rm fm}^3,
\end{equation}
to be compared with $\alpha_p+\beta_p=14.0\pm 0.3$
obtained from photon absorption data through the Baldin
sum rule \cite{levchuk00}. 
A corresponding experiment for the neutron is extremely
difficult, because  there is  no Thomson amplitude to
interfere with the non-Born part of the scattering 
amplitude what provides  the largest contribution to the $\alpha$ and $\beta$
dependent part of the differential
cross section in case of the proton. Nevertheless,
Compton scattering by the neutron at energies below $\pi$ threshold
has successfully been carried out and has led to a first
result for $\alpha_n-\beta_n$ with a meaningful precision
\cite{rose90}. 

Because of the difficulties mentioned above it has been
proposed \cite{levchuk86} to measure quasi-free Compton scattering 
by the neutron at energies between pion threshold and the $\Delta$ 
peak  to determine $\alpha_n -\beta_n$.
This is possible with high precision because free and quasi-free
Compton scattering experiments by the proton have shown, that the
prediction of the theory of quasi-free reactions is valid
\cite{wissmann99}. Also, the prediction of dispersion theory for 
the backward spin polarizability of the proton proved to be valid 
\cite{camen02}. 
Therefore, the corresponding prediction for the neutron, $viz.$
$\gamma^{(n)}_\pi= 58.6\cdot 10^{-4}{\rm fm}^4$,
may be applied to evaluate quasi-free Compton scattering by the neutron.
Recent  investigations \cite{kossert02} have been carried out
with the Mainz 64 cm $\oslash$ $\times$ 48 cm NaI(Tl) detector 
operating in coincidence with the G\"ottingen segmented recoil detector
SENECA by using  deuterium and  hydrogen targets.
Since  quasi-free events from  the proton and neutron are registered
simultaneously and are discriminated  through a thin plastic scintillator 
layer in front of SENECA further tests of the experimental method 
were possible by evaluating the quasi-free proton data in the same 
way as the neutron
data and comparing the results with the corresponding free-proton data.
The result obtained for the polarizability difference of the neutron is
\cite{kossert02}
\begin{equation}
\quad\quad
\alpha_n - \beta_n = 9.8 \pm 3.6({\rm stat}){}^{+2.1}_{-1.1}({\rm syst})
\pm 2.2({\rm model}),\label{alphan}
\end{equation} to be compared with $\alpha_n+\beta_n= 15.2\pm 0.5$
obtained from the photoabsorption cross section \cite{levchuk00}.
A summary of our recent experimental results  is given in Table 1. 
It should be noted 
that the number given for $\gamma^{(n)}_\pi$ is  the first experimental 
result obtained for this quantity, which appears to be slightly 
larger than the previous sum-rule 
prediction \cite{lvov99}. The present result for $\alpha_n$ 
is  in agreement with the result of an experiment on electromagnetic
scattering of neutrons in a Coulomb field \cite{schmiedmayer91}. A
difference, however, of the two methods is that quasi-free Compton
scattering is well tested through experiments on the proton  whereas 
electromagnetic scattering may have large uncertainties \cite{enic97}.

\end{document}